\documentclass[preprint2]{aastex}

\newcommand\hour{\mbox{$^{\mathrm h}$}}
\newcommand\minutes{\mbox{$^{\mathrm m}$}}

\begin{document}

\title{Doppler Modulation of X-ray lines in Cygnus X-3}

\author{Michael J. Stark and Malinda Saia}
\affil{Department of Physics, Lafayette College,
    Easton, PA  18042}
\email{starkm@lafayette.edu, saiam@lafayette.edu}

\begin{abstract}

We measured Doppler shifts of three bright spectral lines in the X-ray
emission from Cygnus X-3 as recorded by the {\it Chandra X-Ray
Observatory\/}.  Doppler shifts of lines associated with \ion{Si}{14}
and \ion{S}{16} exhibit orbital modulation.  The magnitude and phasing
of this modulation relative to the orbital ephemeris indicate the
location of the source of this emission within the wind emanating from
the compact object's companion.  These observations enable us to make
an indirect measurement of the separation of the two stars.  Under
certain assumptions our observation of a line associated with
\ion{Fe}{25} also limits the mass of the compact object $M_{\rm C}\le
3.6M_\Sun$.

\end{abstract}


\section{Introduction}

In spite of near continuous study at many wavelengths since its
discovery \citep{gia67}, the nature of the Cygnus X-3 binary system
remains mysterious.  Several important questions remain about the
nature of the compact object and its companion and about the specific
location of the source of observed X-ray and infrared emission.
Existing models describe the system either as a high-mass system
consisting of a black hole and a Wolf-Rayet star \citep{che94} or as a
low-mass system consisting of a neutron star and a degenerate
companion \citep{tav89}.  Both of these models account for the intense
X-ray emission, the existence of a Helium-rich stellar wind, and the
resulting large mass loss from the system.

The observation of a 12.6~ms pulsar signal in gamma ray data
\citep{cha85} is evidence that the compact object is a neutron star.
However, though this pulsar signal has been repeatedly observed by the
group that first reported it \citep{bow92}, it has not been confirmed
by independent investigators and it has never been observed at lower
energies.  Non-observation at lower energies is not incompatible with
observation at higher energies since the scattering distance for
X-rays in the dense wind is much shorter than for gamma rays.  If
true, the pulsar detection is proof that the compact object is a
neutron star.  Absent independent confirmation, however, the reports
of a pulsar signal must be regarded with skepticism.

One key to understanding the nature of the two stars in the system
would be a direct measurement of their masses.  By measuring the
Doppler shift of the infrared emission, \citet{sch96} report that they
have measured the velocity of the infrared companion and constrained
the compact object to be, at the least, a very massive neutron star.
Hanson, Still, \& Fender (2000) use the Doppler shift of a different
infrared feature to make a more correct measurement of the velocity of
the companion which prefers lower mass systems except that it also
implies that the companion must be the dominant mass in the system.
This rules out systems consisting of a neutron star with a white dwarf
companion.  Knowing the velocity of only one of the stars of a binary
system does not constrain the total mass of the system.  To determine
the masses of both stars, we must make a measurement of the
velocity of the compact object.

\section{Observations}

We used data collected by the {\it Chandra X-Ray Observatory\/} ({\it
CXO\/}) during the radio quiet period prior to the April 2000 radio
outburst \citep{mcc00}.  Data were taken on two occasions two days
apart and each observation lasted approximately five hours, completely
covering one orbital period of the source.  The observations were
carried out using the High Energy Transmission Grating Spectrometer
(HETGS).  Due to photon pileup in the Advanced CCD Imaging
Spectrometer caused by the brightness of Cygnus X-3, the automatic
processing was unable to pinpoint the center of the zeroth order
maximum of the grating spectra.  We refined the automatically
determined source location by fitting prominent spectral features in
both the positive and negative first order spectra of the Medium
Energy Grating (MEG) and the High Energy Grating (HEG).  The source
location was adjusted so that the fitted wavelengths for all four
first order spectra were identical.  By this method, the location of
the X-ray source was determined to be 20\hour 32\minutes 25\fs
8$\pm$0\fs 1 $+$40\arcdeg 57\arcmin 28\farcs 0$\pm$0\farcs1 which,
given the 0\farcs6 absolute astrometric accuracy of the {\it CXO \/},
is consistent with the radio location given by \citet{ogl01}.

We divided the data into eight bins by the orbital phase of the source
according to the ephemeris in Table \ref{ephemtable}.  In each of
these phase bins, we fit a continuum model and Gaussian line profiles
to data from both positive and negative first-order spectra from the
HEG and the MEG simultaneously.  The continuum was fit by a power law
spectrum with a model of interstellar absorption.  Though the distance
to Cygnus X-3 is still poorly determined, the degree of the absorption
fit by the spectrum, $~8.5\pm0.1\times 10^{22}$~\ion{H}{1}/cm$^{2}$ is
consistent with other measurements of the distance to Cygnus X-3 or of
the intervening absorption column \citep{ser75, dic83, pre00}. Our
measured value is also independent of orbital phase.


\begin{deluxetable}{r@{}l}
\tablecaption{Cygnus X-3 orbital ephemeris\label{ephemtable}}
\tablehead{\multicolumn{2}{c}{$T_n=T_0+P_0n+P_0\dot Pn^2/2$}}
\tablewidth{0pt} \startdata $T_0=$&JD $2440949.89016\pm 0.00064$\\
$P_0=$&$0.19968462\pm 6\times 10^{-8}$ d\\ $\dot P=$&$(5.52\pm
0.12)\times 10^{-10}$ \enddata \tablecomments{$T_n$ are the times of
successive X-ray minima from Stark \& Saia, in preparation.}
\end{deluxetable}

\begin{figure}
\plotone{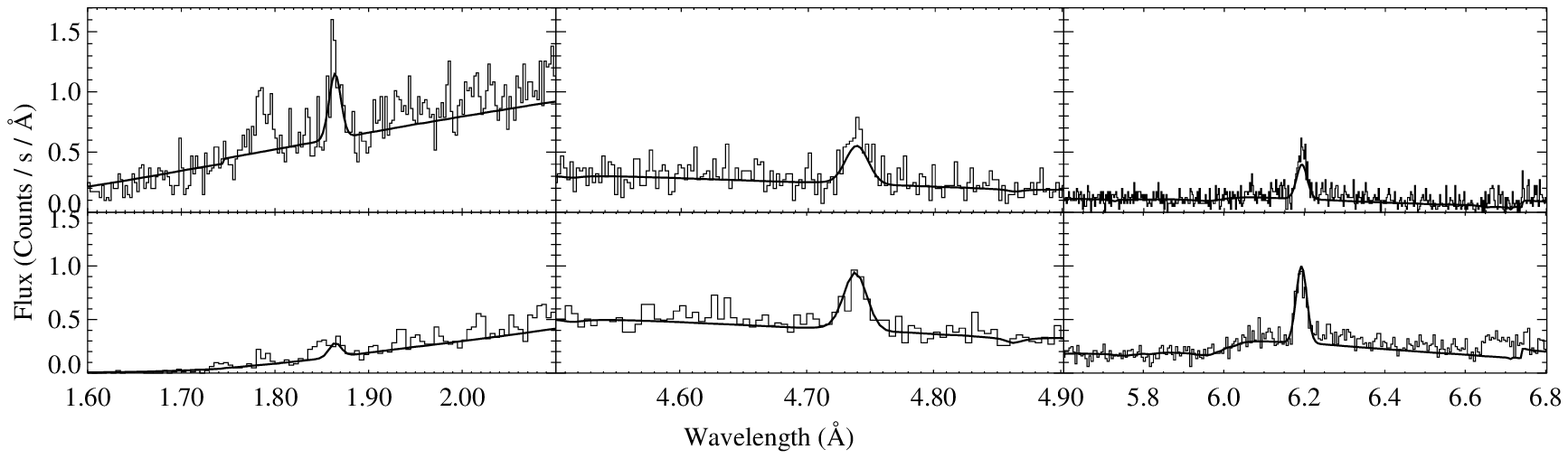}
\caption{Examples of line fitting from this analysis.  The fits
indicated here by the thick line are for the phase interval centered
on 0.625 on 2000 April 4, The left panels are for \ion{FE}{25}, the
center panels are for \ion{S}{16} and the right panels are for
\ion{Si}{14}.  The top panels are from the positive first order HEG
spectrum and the bottom panels are from the positive first-order MEG
spectrum.  The fitted profiles are derived from fitting all four of
the first-order spectra simultaneously. \label{samplines}}
\end{figure}

We fit a Gaussian line function to the three most prominent lines
visible in the spectrum, the Lyman $\alpha$ lines of \ion{Si}{14} and
\ion{S}{16} and the Helium $\alpha$ line of \ion{Fe}{25}.  Only these
three lines were strong enough in the phase-dependent spectra that
they could be fit reliably for all phase intervals.  Examples of these
fits are shown in Figure \ref{samplines}.  The \ion{Si}{14} and
\ion{S}{16} lines showed a consistent average Doppler shift
corresponding to a recession velocity of $\sim 550$ km/s.  This is
somewhat less than the velocity of 750--800~km/s reported by
\citet{pae00}, also from HETGS data, but nearly consistent given the
absolute spectral accuracy of the HETGS is on the order of 100~km/s.


\begin{figure}
\plotone{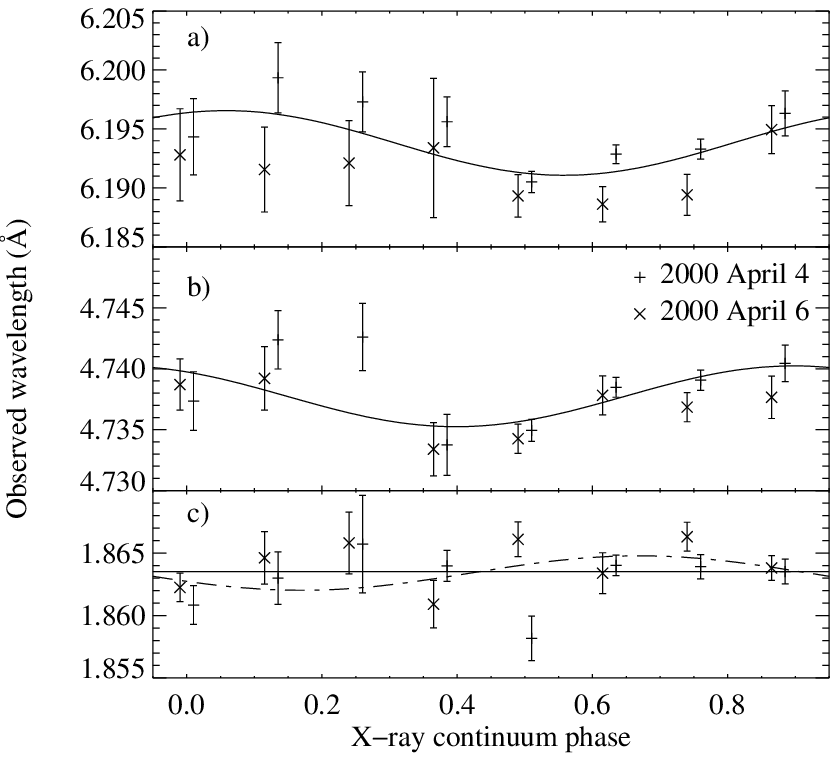}
\caption{Observed wavelength of (a) the \ion{Si}{14} Ly~$\alpha$ line
and (b) the \ion{S}{16} Ly~$\alpha$ line and (c) the \ion{Fe}{25}
He~$\alpha$ line as a function of the X-ray phase. The \ion{S}{16}
point at phase 0.25 from 2000 April 6 is corrupted by a peculiarity of
the background and is excluded from the fit.  The phases on the plots
have been shifted slightly to reveal overlapping data points.  The
dashed line in (c) corresponds to the 90\% confidence upper limit on
the modulation of the \ion{Fe}{25} line.\label{shift}}
\end{figure}

Figure \ref{shift}(a) shows the Doppler shift of the 6.18~\AA\ Lyman
$\alpha$ line of \ion{Si}{14} and Figure \ref{shift}(b) shows the
Doppler shift of the 4.73~\AA\ Lyman $\alpha$ line of \ion{S}{16}.
The parameters of the best fit of a cosine function to the Doppler
shift of each line are given in Table \ref{dopfit}.  The sinusoidal
function provides a better fit to the data than does a constant
function.  For \ion{Si}{14}, $\chi^2$ per degree of freedom drops from
2.7 to 1.6 with the change from a constant function to a
sinusoidal function.  In \ion{S}{16} the change in $\chi^2$ per degree
of freedom is from 2.8 to 1.5.  The F-test probabilities for these
detections are 18\% and 14\%, respectively, so the chance probability
of the simultaneous detection of these two modulations is 2.5\%.


The Doppler shift of the \ion{Si}{14} line has significant day-to-day
variation though the modulation is clearly present on both days.  The
modulation of the \ion{S}{16} line appears to be particularly
repeatable.  Sinusoidal functions may not be the most appropriate
functional forms to fit to these data.  In the simplest configuration,
the emission region will be in the form of a ring or a disk near the
compact object sampling different wind velocities. In addition,
different parts of the emission region may be occulted at different
times.  Beyond the apparent inadequacy of the shape of the fit
function, the phase of the maximum Doppler shift may be affected by
the differential occultation of the emission region.  Nevertheless,
the amplitude of the fit sinusoid gives a reasonable approximation of
the amplitude of the correct functional form.  The quality of
these fits to the data ($\chi^2$ per degree of freedom is 1.6 for
\ion{Si}{14} and 1.5 for \ion{S}{16}) indicate that within our
uncertainties the true functional form is not much more complicated
than a sinusoid.


\begin{deluxetable}{lclcllr}
\tablecaption{Doppler Shift of Strong Emission Lines\label{dopfit}}
\tablehead{& \colhead{Source} & \colhead{Observed} &
\colhead{Recession} & \colhead{Modulation} &
\colhead{Modulation}&\colhead{Phase}\\ \colhead{Line} &
\colhead{Wavelength} & \colhead{Wavelength} &
\colhead{Velocity\tablenotemark{a}}&\colhead{Amplitude}&\colhead{Amplitude}&\colhead{Offset\tablenotemark{b}}\\
&\colhead{(\AA)}&\colhead{(\AA)}&\colhead{(km/s)}&\colhead{(\AA)}&\colhead{(km/s)}&\colhead{($^\circ$)}} \tablewidth{0pt} \startdata
\ion{Si}{14}&$6.1822$&$6.19381\pm0.00058$&$563\pm28$&$0.0027\pm0.0007$&$133\pm35$&$21\pm14$\\
\ion{S}{16}&$4.7292$&$4.73774\pm0.00045$&$541\pm29$&$0.0025\pm0.0006$&$158\pm35$&$-37\pm14$\\
\ion{Fe}{25}&$1.8617$\tablenotemark{c}&$1.86340\pm0.00038$&$274\pm61$\tablenotemark{c}&$<0.0014$\tablenotemark{d}&$<220$\tablenotemark{d}&\nodata
\enddata
\tablenotetext{a}{Quoted uncertainty is from the fit of a Gaussian
line profile.  Absolute spectral accuracy of the HETGS is on the order of
100~km~s$^{-1}$.}
\tablenotetext{b}{Phase offset is for the maximum red-shift relative
to the time of the minimum intensity of the X-ray continuum.}
\tablenotetext{c}{The reported \ion{Fe}{25} source wavelength is from
a combination of lines we may have modeled incorrectly.  The value of
the recession velocity is, therefore, subject to a large, unknown
systematic error.  It is not, however, important to the results
presented here.}
\tablenotetext{d}{Upper limits are 90\% confidence limits.  All other
uncertainties are $1\sigma$}
\end{deluxetable}


Figure \ref{shift}(c) shows the Doppler shift of the 1.86~\AA\
\ion{Fe}{25} He~$\alpha$ line.  The data are consistent with a constant
Doppler shift.  The $\chi^2$ per degree of freedom assuming a constant
Doppler shift is 1.77 while the best fit sinusoidal function yields a
$\chi^2$ per degree of freedom of 1.84.  The non-observation of a
modulation of the Doppler shift in \ion{Fe}{25} allows us to set an
upper limit on the velocity of the emission region for this line.  The
90\% confidence upper limit derived for a fit to these data is
220~km/s.

\section{Discussion}

In order to use our observations to draw conclusions about the size
and mass of the Cygnus X-3 system, we must first decide where in the
system the emission is produced.  \citet{vke93} observed that line
emission from hydrogen-like helium was produced in that part of the
stellar wind of the companion which is shaded by the companion from
the X-rays emanating from the compact object. According to his model,
the rest of the wind is too highly ionized to produce helium line
emission.  The coincidence between the maximum red shift of the line
emission from hydrogen-like silicon and sulfur and the time of the
X-ray minimum (Taken to indicate the superior conjunction of the
compact object) suggests that these emission features are produced in
the companion wind near the compact object.  The fact that the silicon
and sulfur emission are produced much nearer to the X-ray source than
the hydrogen emission suggests that the dominant source of ionization
in the Cygnus X-3 system is the X-rays.  The level of ionization
appears to increase with proximity to the source of X-rays. Emission
from hydrogen-like and helium-like iron will thus be produced even
closer to the X-ray source---presumably the compact object---than is
the silicon or sulfur emission.  We will, therefore, assume that the
emission feature produced by helium-like iron is produced close to
the compact object.

\subsection{Stellar Mass}

The emission features of highly ionized iron may be produced in a
region of the wind captured by the compact object or from an accretion
disk around the object if not from the surface of the object itself.
Without knowing the phase relationship between the Doppler shift of
the \ion{Fe}{25} line and the X-ray ephemeris, we can not be sure that
it would reveal the motion of the compact object.  If, however, we
assume that this is so we can use our non-observation of Doppler shift
in the \ion{Fe}{25} line to constrain the velocity of the compact
object.  The equation for the mass function
\begin{equation}
m_f={1\over2\pi G}Pv_{\rm Max}^3 \left( 1-\epsilon^2\right)^{3/2}
\label{massfunc}
\end{equation}
yields an upper limit on the mass function of 0.22~$M_\sun$ taking
$P=4.8$~h and our 90\% confidence upper limit of 220~km/s for the
maximum radial velocity, $v_{\rm Max}$.  \citet{gho81} show that a
high mass-loss rate can drive the system into an eccentric orbit as
well as cause it to circularize.  For simplicity we will assume
$\epsilon=0$ for the time being.

\begin{figure}
\plotone{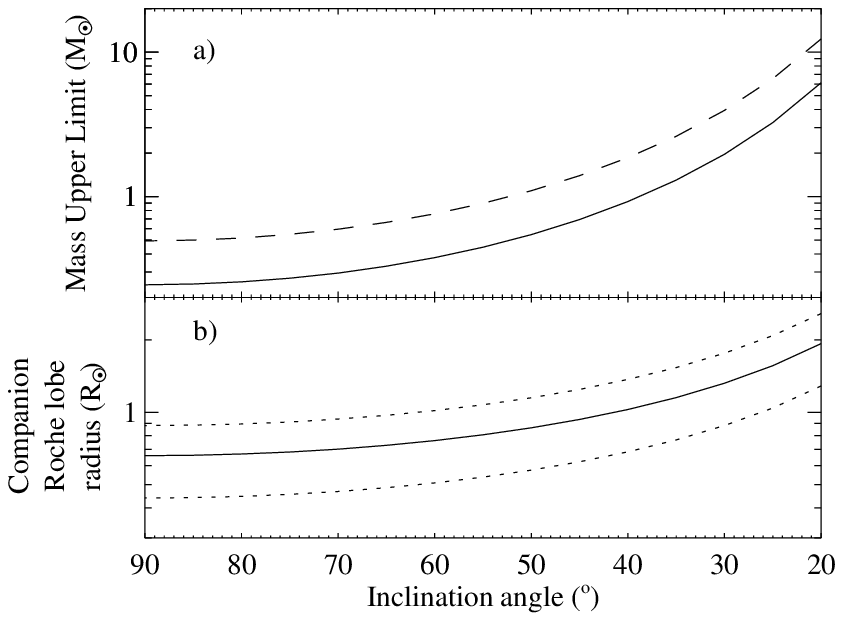}
\caption{(a) Mass upper limits of the compact object (solid line) and
the companion (dashed line) and (b) the Roche lobe radius of the
companion (dotted lines are 1$\sigma$ limits) in the Cygnus X-3 system
as a function of inclination angle.\label{orbit}}
\end{figure}

By measuring the Doppler shift of the centroid of an infrared
absorption feature in the stellar wind \citet{han00} calculated a mass
function for the wind producing companion of 0.027~$M_\Sun$.  We can
combine this mass function with our mass function upper limit by
simultaneously solving the following equations
\begin{equation}
{m_f}_{\rm D} = {\left( M_{\rm C} \sin i\right)^3 \over \left( M_{\rm
D} + M_{\rm C}\right)^2} , {m_f}_{\rm C} = {\left( M_{\rm D} \sin
i\right)^3 \over \left( M_{\rm C} + M_{\rm D}\right)^2}
\label{masses}
\end{equation}
Where the subscript C refers to the compact object and the subscript D
refers to the mass donating companion.  $i$ is the inclination of the
binary orbit.  The upper limits on the masses of the two stars derived
this way are $M_{\rm C} \le 0.24 M_\Sun$ and $M_{\rm D} \le 0.49
M_\Sun$ for an inclination, $i=90^\circ$.  These limits essentially
exclude all stars that could produce the observed X-ray flux.  Mass
upper limits as a function of orbital inclinations are shown in Figure
\ref{orbit}(a).


The orbital model of Cygnus X-3 that has been used most extensively is
one of two stellar wind models proposed by \citet{gho81} which have a
large orbital inclination and a large eccentricity.  Even
\citet{han00} state that the inclination should be large, but our
results, taken in combination with theirs, make large inclinations
unphysical.  In light of other recent work \citep{sin02} it appears
that the lack of apsidal motion in the system requires that the
orbital eccentricity be small.  In addition, several authors
\citep{van89, mio01} demonstrate that the orbital inclination of the
system must also be small.  The best fit to the \citet{gho81} model
based on a variable luminosity X-ray source inside its companion's
stellar wind satisfies the need for both a small orbital inclination
($i= 24^\circ$) and eccentricity ($\epsilon = 0.14$).  Using these
values in Equations \ref{massfunc} \& \ref{masses} gives an upper
limit for the mass of the compact object, $M_{\rm C}\le 3.6M_\Sun$ and
for the companion of $M_{\rm D}\le 7.3M_\Sun$.  These mass limits
don't preclude the possibility of the system being composed of a black
hole and a Wolf-Rayet star.  Neither do they rule out the compact
object being a neutron star but the suggestion that its companion is
more massive than the compact object rules out the possibility that
the companion is also degenerate (a white dwarf).

\subsection{Stellar Size}

Table \ref{dopfit} shows that the maximum red shift of the
\ion{Si}{14} and \ion{S}{16} emission lines are nearly coincident with
the minimum of the continuum X-ray intensity.  The coincidence of the
maximum radial velocity of the line emission and the X-ray minimum
means that the line emitting region is moving most directly away from
Earth at the same time that the compact object is at its furthest
distance from Earth.  This, then, suggests that the line emission
emanates from a region of the companion's stellar wind blowing by the
compact object and close to it.  We can therefore use our
measured velocity for this wind with a model of wind acceleration to
constrain the size of the binary system.

The standard model of stellar wind acceleration appropriate to
Wolf-Rayet stars is
\begin{equation}
v(r)=v_\infty\left( 1-{r_c\over r} \right)^\beta
\end{equation}
where $r_c$ is the core radius of the star and $\beta$ expresses the
degree of acceleration.  Alternately, the wind velocity can be
expressed in terms of the extent of the accelerating wind,
\begin{equation}
v(r)=v_\infty \left( r\over r_\infty \right)^\gamma
\end{equation}
\citep{che94} where $r_\infty$ is the distance from the wind producing
star to where the terminal velocity is reached.  We can combine the
wind velocities implied by the analysis of the \ion{Si}{14} and
\ion{S}{16} lines to provide an estimate of the wind velocity in the
emission region surrounding the compact object of $146\pm50$~km/s.
Taking $v_\infty = 1500$~km/s \citep{sch96}, $r_\infty =15R_\Sun$
\citep{fen99}, and $\gamma =1$ \citep{ant01}, the distance between the
emitting region and the wind producing companion star is
$1.5\pm0.5R_\Sun$.  This result assumes an inclination $i=90^\circ$.
For an inclination angle $i=24^\circ$, we calculate $3.6\pm1.2R_\Sun$
for the distance between the donor star and the emission region.
Since this emission comes from near the X-ray star, the distance
calculated between the emission region and the companion star is the
distance between the two stars.

The size of the Roche lobe of each star can be calculated directly
from the mass functions.  The radius of the Roche lobes
is given by
\begin{equation}
{r_{\mathrm L}\over a}={0.49q^{2/3}\over 0.6q^{2/3}+\ln \left( 1+q^{1/3}\right) }
\end{equation}
\citep{egg83} where $a$ is the binary separation and $q$ is the ratio
of the stellar masses, which is only dependent on the mass functions
and which in the case of the results presented here $M_{\rm D}/M_{\rm
C}=2.0$.  This yields a Roche lobe radius of the companion star of
$R_{\rm D}=1.6\pm0.5R_\Sun$ at an inclination, $i=24^\circ$, which is
somewhat small for a Wolf-Rayet star \citep{mof96} especially given
that the dominant mass-transfer process is the stellar wind rather
than Roche Lobe overflow.  The companion Roche lobe radius is plotted
as a function of the inclination of the system in Figure
\ref{orbit}(b).

\section{Conclusion}

We have determined an upper limit on the radial motion of the X-ray
emitting star.  This limit assumes that the Doppler shift of the
emission feature produced by \ion{Fe}{25} would reveal the motion of
the X-ray emitting star.  The relative phases of the observed Doppler
shift of the \ion{He}{2}, \ion{Si}{14}, and \ion{S}{16} emission
features suggest that the degree of ionization of the circumstellar
material is correlated with proximity to the X-ray source.  This is
consistent with our assumption.  Combined with several other
observations, we begin to paint a consistent picture of the Cygnus X-3
system.  The picture favored by our analysis is of a system consisting
of two relatively low mass stars, a compact object $M_{\rm C} \le
3.6M_\Sun$ and its companion $M_{\rm D} \le 7.3M_\Sun$ which orbit
each other separated by $3.6\pm1.2R_\Sun$ in relatively circular
$\epsilon=0.14$ orbit that has a small inclination $i=24^\circ$
relative to our line of site.  Future X-ray observations will enable
us to test the assumptions underlying this picture and, if they prove
correct, to determine the orbital parameters of the system.

\end{document}